\documentclass{article}
\usepackage{spconf,amsmath,graphicx}
\usepackage{multirow}
\usepackage[table,xcdraw]{xcolor}
\usepackage{pdfpages}
\usepackage{pifont}
\usepackage{booktabs}
\usepackage{multirow}
\usepackage{caption}
\usepackage{subcaption}

\title{ADAPTING LEARNED IMAGE CODECS TO SCREEN CONTENT VIA ADJUSTABLE TRANSFORMATIONS }
\name{\begin{tabular}{c}H. Burak Dogaroglu$^{1}$ \qquad A. Burakhan Koyuncu$^{1,2}$ \qquad Atanas Boev$^{2}$ \\ Elena Alshina$^{2}$ \qquad Eckehard Steinbach$^{1,3}$\end{tabular}}

  \address{$^{1}$ School of Computation, Information and Technology, TU Munich \\
    $^{2}$ Huawei Munich Research Center \\
    $^{3}$ Munich Institute of Robotics and Machine Intelligence \\
    Email: burak.dogaroglu@tum.de} 

\begin{document}
%
\maketitle
\begin{abstract}
As learned image codecs (LICs) become more prevalent, their low coding efficiency for out-of-distribution data becomes a bottleneck for some applications. To improve the performance of LICs for screen content (SC) images without breaking backwards compatibility, we propose to introduce parameterized and invertible linear transformations into the coding pipeline without changing the underlying baseline codec's operation flow. We design two neural networks to act as prefilters and postfilters in our setup to increase the coding efficiency and help with the recovery from coding artifacts. Our end-to-end trained solution achieves up to 10\% bitrate savings on SC compression compared to the baseline LICs while introducing only 1\% extra parameters.
\end{abstract}
\begin{keywords}
screen content coding, learned image compression, prefilter, postfilter, deep learning
\end{keywords}
\section{Introduction}
\label{sec:intro}

As available compute power of consumer grade hardware increases, artificial intelligence (AI) products start to replace more classical solutions. Image compression is one of these fields where the neural network-based learned image codecs (LICs) \cite{balle2018variational,minnen2018joint,cheng2020learned,koyuncu2022contextformer,koyuncu2023efficient,he2022elic,liu2023learned} can produce similar level or more efficient encodings than their classical counterparts. This opens up opportunities to develop more efficient AI-based coding standards for image and video compression.

Screen content (SC) images have unique characteristics. Unlike natural images where the pixels are produced by sensors capturing natural phenomena such as light, the SC pixels are mostly artificially generated. This causes SC to have drastically different color and texture patterns \cite{Wang2022}. They usually consist of sharp edges with reduced colorspaces. Some examples for SC images are graphical user interfaces, large bodies of letters and text, computer simulation results and so on. It is also possible to have natural images embedded in a SC context, e.g. game streaming where the user's video feed is combined with artificially generated game images.

The SC compression is often overlooked in the domain of LICs due to lack of representation in most datasets and lack of effort in the exploitation of unique redundancies. It is also very hard to update LICs after they are standardized since any change in network weights may break the backwards compatibility of the bitstreams and it's not feasible to store multiple copies of large networks for every iteration. For these reasons, we research solutions in adapting LICs to out-of-distribution data like SC without breaking the compatibility of already generated encodings.

In that regard, we develop new modules around the codec that can be disabled with a signal. This setup allows us to improve coding efficiency of LICs without retraining the codec itself. For this purpose, we develop two neural networks that perform as the prefilter and postfilter around the codec's input and output. To enhance the performance on SC compression, we also propose the use of linear forward and inverse transformations.

The prefilter model focuses on creating better compressible images through addition of spatial modulations to the source image. On the other hand, the postfilter model aims to remove the modulations and reduce the artifacts created by the forward transformation and the prefilter model. We train our models end-to-end so that they can learn to use information transmitted through the codec more efficiently.




\begin{figure*}[t]
  \centering
  \includegraphics[width=0.9\textwidth]{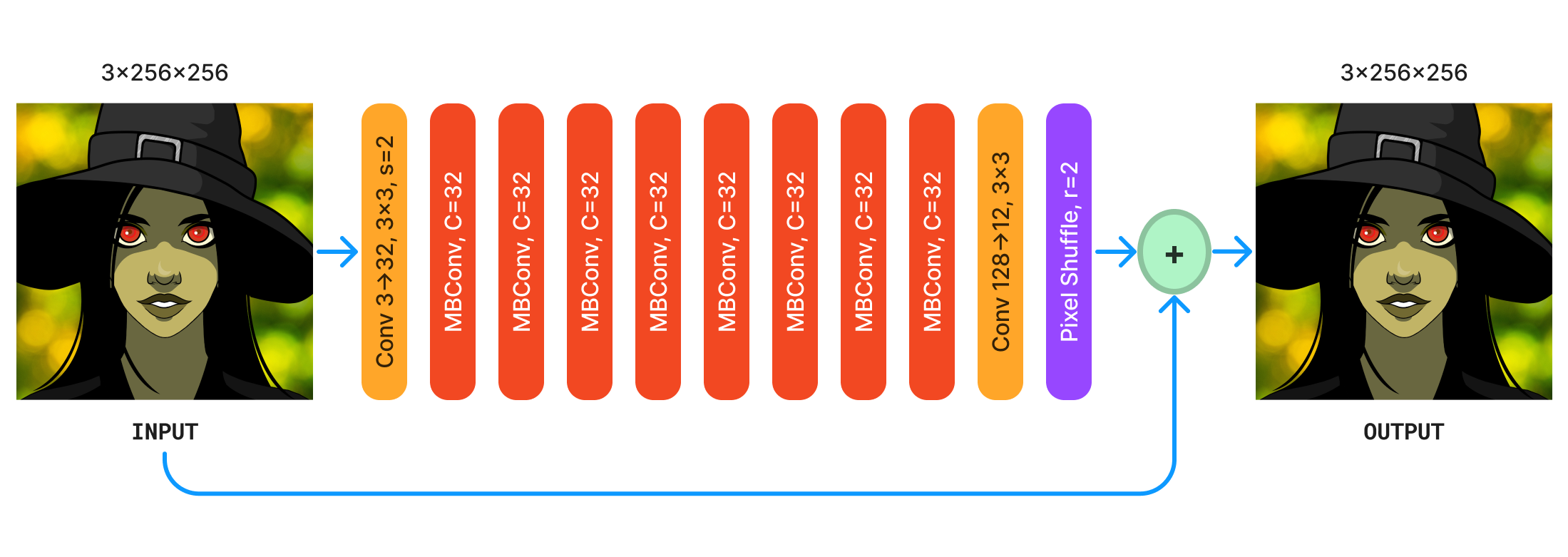}
\caption{Architecture of the residual network based on MBConv layers. This implementation depicts $L = 8, C=32$ setup.}
\label{fig:resnet}
\end{figure*}

\section{Background}
\label{sec:background}

\subsection{Classical screen content compression}
Rise of digitalization has increased the demand for SC compression. As a response, video codec standards such as HEVC \cite{hevc} and VVC \cite{vvc} adopted specialized extensions. Among these, most prominent ones are the Intra Block Matching (IBM) and Palette Mode Coding (PMC) \cite{liu2015overview,overview_recent_video_scc}.

In IBM, for each coding unit, the intra encoder searches for similar or repeating patches in the already encoded parts of the image. This allows the encoder to skip encoding large chunks of pixels and just transmit the position of the similar patch to the decoder. In turn, the decoder just copies the referenced block from already decoded coding units. This technique increases the coding efficiency especially when the source image has large chunks of repeating blocks, such as graphical user interfaces, pixel art or gaming.

On the other hand, PMC focuses on exploiting the color space redundancy of SC images. The algorithm encodes the set of unique colors in a given block and replaces the pixels with the indexes corresponding to their colors. Instead of transmitting the color channels of the image, we encode the palette and send the indexes alongside it. This way, we can get lower bitrates for source images with narrow color spaces.

\subsection{Learned image compression}
After the invention of differentiable entropy bottleneck layers and end-to-end trainable autoencoder codecs \cite{balle2016end}, learned image compression has seen growing interest from the researchers. Later, hyper-encoders and hyper-decoders were introduced to decrease the spatial redundancy on latent codes \cite{balle2018variational}. This was further improved by allowing a context model to process the reconstructed latents \cite{minnen2018joint}. Designing context models with better priors were shown to reduce required bitrate to transmit images \cite{cheng2020learned}. Researchers began to test transformer based context models to replace the autoregressive architectures \cite{koyuncu2022contextformer}. Since transformer context models required more compute, there has been also some attempts to make them more efficient via routing attention to selected channels \cite{koyuncu2023efficient}.

\subsection{Hybrid approaches}

Tao et al. \cite{comcnn} combined two neural network models around a classical image codec and trained them together to improve the coding efficiency of the underlying codec. To get around the non-differentiable codec's lack of backpropagation, they used auxilary loss functions to train the networks separately. Recently, Guleryuz et al. \cite{sandwich1} solved the same problem by switching the actual codec with a differentiable image proxy during training time. This idea was also combined with applying transformations such as downscaling or quantization to enhance codec capabilities \cite{sandwich2}.

\section{Methodology}
\label{sec:methodology}
Training a LIC is computationally expensive. In this work, we attempt to improve the SCC efficiency of an existing LIC without retraining of the whole codec, but using pre- and post-processing operations instead. The codec is treated as a black box and its weights are frozen during training. This helps our solution to be implemented easily on top of any differentiable codec while keeping the produced bitstream compatible with the underlying software. 

\subsection{Proposed solution}

\begin{figure}[!t]
\centering
\includegraphics[width=0.9\linewidth]{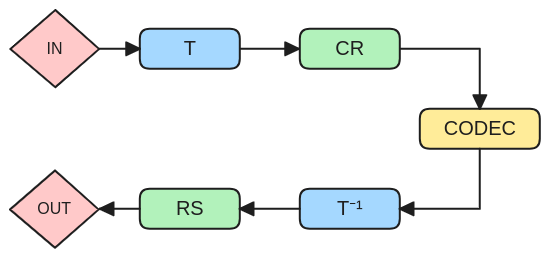}
%
\caption{Operational diagram of the proposed pipeline.}
\label{fig:pipeline}
\end{figure}

Our solution requires four extra modules on top of the codec. In the encoder side, we have a linear forward transformation $T$ and Compact Representation (CR) module with a neural network. Following a similar structure, the decoder consists of a linear inverse transformation $T^{-1}$ and the Reconstruction Stage (RS) module. Fig.~\ref{fig:pipeline} depicts the operational flow of this solution.

\subsection{Linear transformations}

In our setup, we can't change the internal flow of the codec. That means, we must work with the same kind of inputs and outputs as the codec. This puts a requirement on the valid transformations such that they must be compatible with 3-channel RGB images. In cases where the exact inverse transformation is not possible, we can approximate it with the Moore - Penrose inverse \cite{Penrose1955} of the forward transformation.

In this work, we consider 3 candidate transformations. Desaturation transformation takes an RGB image and linearly interpolates between the grayscale version according to the formula $ \hat{x} = x + (1 - \alpha) Gray(x) $ where $x$ is the RGB input, $Gray$ is the 3-channel grayscale color conversion and $\alpha$ is the saturation level. Its inverse can easily be computed by just changing $\alpha$ to its multiplicative inverse.

The PCA downscaling transformation first converts an image to its PCA colorspace consisting of the principal channel (PC), side channel 1 (SC1) and side channel 2 (SC2) according to the order of lowest standard deviation. Then, it downscales the side channels by the parameters $d_{SC1}$ and $d_{SC2}$ respectively where the downscaled images have image dimensions divided by these values. While bicubic transformation is used during downscaling, we preferred to upscale using the nearest neighbor approximation so the color channels become the same resolution as the original image. Once the image is constructed back, we remap the colors to RGB using the inverse PCA color transformation. 

Similarly, we also evaluate PCA quantization transform which uses the same color transformation as before. However, instead of downscaling, we quantize the side channel amplitudes with $Q_{SC1}$ and $Q_{SC2}$ bits by replacing them with cluster centers produced by the k-means \cite{Lloyd1982} algorithm. Finally, the quantized image is remapped back to the RGB space to be consumed by the LICs. 

\subsection{Network architecture}

Our CR and RS modules use the exact same architecture with same number of layers and channels. We use a ResNet architecture~\cite{He2016} where the main processing blocks are MBConv layers~\cite{effnet}. The network consists of a convolution layer with stride of 2 that maps RGB images to $C$ channels, $L$ number of MBConv layers where the last layer doesn't squeeze its channels, another convolution layer that maps the unsqueezed $4C$ channels to 12 so that we can finish with pixel shuffling \cite{shi2016real} with a stride of 2 to get the original resolution again. To stabilize the training and let network focus on only extracting modulations, we have an end-to-end residual connection between the input and output of the module. This structure is illustrated in Fig.~\ref{fig:resnet}.

\subsection{Dataset}

We used the SC dataset collected for the JPEGAI \cite{ascenso2023jpeg} standardization effort, denoted as JPEGAI-SC hereinafter. It has approximately 3 thousand training images and 1 thousand validation images. We use the validation set in our analysis section since we don't use it in any part of training. The dataset consists of 5 categories of SC images: AI-generated, gaming, screenshots, 3D renderings, illustrations.

\subsection{Training scheme}

Our models were trained for 50 epochs where the codecs' weights were frozen. The batch size in our trainings was set to 8 and we worked with center cropped 256x256 images both for training and testing. We used Adam \cite{kingma2014adam} optimizer with $\gamma=10^{-4}, \beta_1=0.9, \beta_2=0.999$. The codec implementations are from the CompressAI \cite{begaint2020compressai}. We used the same Rate - Distortion loss with the same $\lambda$ parameters as the baseline codecs \cite{balle2018variational, minnen2018joint, cheng2020learned}. We optimized the Mean Squared Error (MSE) as the codec variants we use.

\section{Experimental Results}
\label{sec:analysis}

We performed a series of experiments to find optimal parameters and settings for various questions. In most cases, we only trained for quality points between 2 and 5 due to limited access to compute resources and 4 quality points are enough to compute Bjontegaard-Delta (BD) rate with Akima interpolation \cite{Bjontegaard2001CalculationOA, herglotz2022beyond}.

\subsection{Compressibility of transformations}

Since we are interested in creating bitrate reductions through reducing the information of the images with transformations, we designed an experiment where the transformed images are passed through a LIC and we measure their compressed sizes. We used the Ballé encoder for quality points between 2 and 5 and measured averaged the bitrate differences $\Delta BPP$ of the augmented images to their baselines. 

\begin{figure*}[htb]
%
\centering
\includegraphics[width=1.0\linewidth]{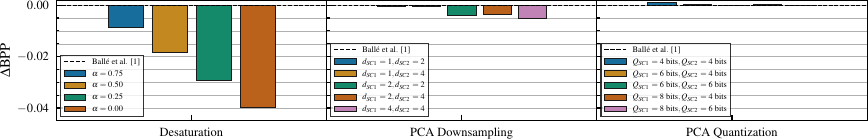}

\caption{Compressibility of desaturation, PCA downsampling and PCA quantization. Lower bitrate indicates a potential for performance gain. While desaturation clearly allows us to compress images into smaller files, the PCA downsampling and PCA quantization transformations have no visible effect on the bitrate difference.}
\label{fig:compressibility}
\end{figure*}

In Fig.~\ref{fig:compressibility}, we can see that while desaturation transformation gets smaller bitrates compared to the Ballé codec, the information loss created by other transformations are not well received by the codec and does not create noticable savings even though they introduce more noise to the system. This indicates that only the desaturation transformation have any potential for performance gains for SC compression, so we focus on only desaturation transformation in the following experiments.

\subsection{Effect of desaturation magnitude}

\begin{figure}[!t]
\centering
\includegraphics[width=1.0\linewidth]{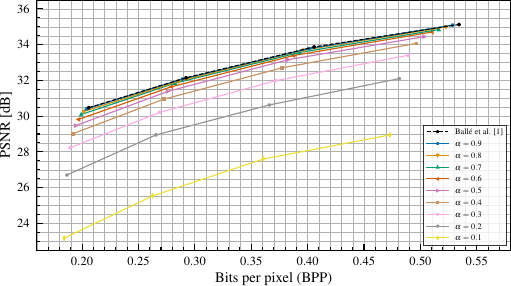}
\caption{Rate-Distortion plot of different desaturation levels on top of Ballé baseline codec. The reconstructed image quality drops higher for stronger transformations. However, weaker transformations produce almost little to none bitrate gains, indicating a sweet spot for the quality - bitrate tradeoff.}
\label{fig:desatlevels}
\end{figure}

We introduced the inverse transformation on the decoder side and measured end-to-end reconstruction performance of the pipeline without the neural network modules. From Fig.~\ref{fig:desatlevels}, it can be seen that the transformations alone hurt the performance more as the saturation level $\alpha$ gets smaller. Assuming our networks can produce 0.5 - 1 dB PSNR gain on top of the codecs, we opt to focus on a saturation level of $\alpha = 0.8$ in the following experiments.

\subsection{Performance analysis}

\begin{figure}[htb]
  \centering
 \includegraphics[width=1.0\linewidth]{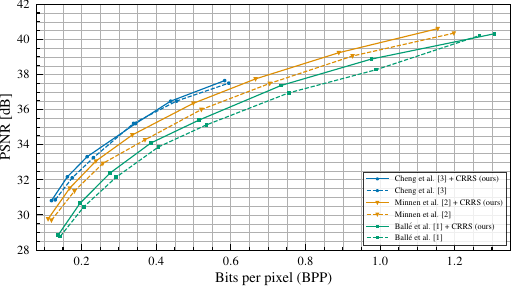}
%
\caption{Rate-Distortion plot for our solution with baselines as Ballé, Minnen and Cheng codecs.}
\label{fig:codecperf}
\end{figure}

Finally, we activate the neural network modules and train the whole pipeline in an end-to-end manner. We repeat this for all available quality points in Ballé et al.~\cite{balle2018variational}, Minnen et al.~\cite{minnen2018joint} and Cheng et al.~\cite{cheng2020learned} baseline codecs. For ease of comparison, we used the same neural network models of $L=8$ layers and $C=32$ channels in all cases. Fig.~\ref{fig:codecperf} shows that in all cases, our modules improved over the baseline codecs.

\subsection{Computational complexity}

We performed an analysis to find out the relative costs and benefits of using our pipeline compared to the baseline codecs. According to Table \ref{table:complexity}, we see that it is possible to adapt learned image codecs to out-of-distribution SC data by using just \%1 more parameters. We also measured the Multiply-Accumulate Operations per pixel (MAC/px) of the end-to-end pipelines and our modules introduce few extra computations to increase the coding efficieny of SC images. We follow the technique proposed in JPEGAI \cite{ascenso2023jpeg} standardization to find the MAC/px of the models. The reported values are from the highest quality point baseline codec available in the CompressAI \cite{begaint2020compressai} library. 

\begin{table}[]
\addtolength{\tabcolsep}{-1pt}
\centering

\begin{tabular}{lccccc}
\toprule
                     & &  & \multicolumn{2}{c}{BD-Rate} \\ 
Model                    & \# Param. & MAC/px & \multicolumn{2}{c}{(\%) $\downarrow$} \\ 
\midrule

Ballé et al.~\cite{balle2018variational} & 11.82M       & 418K                                                         & \phantom{-1}0.0                                                       \\
$\:\:\:$ + CRRS     & 12.02M       & 460K                                                         & \phantom{1}-9.0                                                       \\ \midrule
Minnen et al.~\cite{minnen2018joint} & 25.51M       & 450K                                                         & -20.6 & \phantom{-1}(0.0)                                                  \\
$\:\:\:$+ CRRS   & 25.71M       & 492K                                                         & -28.9 & (-10.5)                                                  \\ \midrule
Cheng et al.~\cite{cheng2020learned} & 26.60M       & 927K                                                         & -36.9 & \phantom{-1}(0.0)                                                   \\
$\:\:\:$+ CRRS    & 26.80M       & 969K                                                           & -39.6 & \phantom{1}(-4.5)     \\                                             
\bottomrule

\end{tabular}
\caption{Model complexity and performance analysis relative to the Ballé baseline codec. For Minnen and Cheng pipelines, relative BD-Rate gains are reported in parantheses. Lower BD-Rate is better. }
\label{table:complexity}
\end{table}

\begin{table}[t]
\addtolength{\tabcolsep}{-1.5pt}

\newcommand{\RomanNumeralCaps}[1]
    {\MakeUppercase{\romannumeral #1}}
\centering
\begin{tabular}{l|ccccc|c|c}
		\toprule
                                                                                     &   &   &   &       &        &  & BD-Rate \\
Name                                                                                    & $\alpha$ & CR  & RS  & $L$       & $C$       & $\Delta$MAC/px &  (\%) $\downarrow$ \\ \midrule
Ballé~\cite{balle2018variational}                              & \ding{55}    & \ding{55} & \ding{55} & \ding{55} & \ding{55} & 0                    & \phantom{-1}0.0                       \\ \midrule
\multicolumn{1}{c|}{\multirow{4}{*}{\RomanNumeralCaps{1}}} & \ding{51}    & \ding{51} & \ding{51} & 5         & 64        & 102K                 & \textbf{-11.5}            \\
                                                                                        & \ding{51}    & \ding{51} & \ding{51} & 3         & 64        & 65K                  & \phantom{1}-9.4                      \\
                                                                                        & \ding{51}    & \ding{51} & \ding{51} & 10        & 32        & 51K                  & \phantom{1}-6.2                      \\
                                                                                        & \ding{51}    & \ding{51} & \ding{51} & 8         & 32        & 42K                  & \phantom{1}-8.0                        \\ \midrule
\multicolumn{1}{c|}{\multirow{5}{*}{\RomanNumeralCaps{2}}}     & \ding{55}    & \ding{51} & \ding{55} & 5         & 64        & 51K                  & \phantom{1}-0.9                      \\
                                                                                        & \ding{55}    & \ding{55} & \ding{51} & 5         & 64        & 51K                  & \phantom{1}-4.4                      \\
                                                                                        & \ding{55}    & \ding{51} & \ding{51} & 5         & 64        & 102K                 & \phantom{1}-7.4                      \\
                                                                                        & \ding{51}    & \ding{55} & \ding{55} & 5         & 64        & 0                    & \phantom{-1}2.0                       \\
                                                                                        & \ding{51}    & \ding{55} & \ding{51} & 5         & 64        & 51K                  & \phantom{1}-2.9                      \\ \bottomrule
\end{tabular}
\caption{Ablation study with Ballé baseline codec on quality points 2-5. We investigate the effect of different model size in (\RomanNumeralCaps{1}) and the effect of each proposed modules such as the desaturation, CR and RS in (\RomanNumeralCaps{2}). We set the desaturation level to 0.8 for these experiments.}
\label{table:ablation}
\end{table}

\subsection{Ablation study}

Our ablation study has two steps. In the first step, we kept every module active and tested the performance of changing number of channels $C$ and number of layer $L$ for our neural network models. In the second step, we fixed the neural network architecture and only measured the performance contribution of modules separately and together. Table \ref{table:ablation} summarizes our findings from these experiments.

From the first stage ablation experiments, in general, increasing the model complexity helps with the performance. However, increasing the number of layers from 8 to 10 for 32 channels is an exception to this observation. We hypothesize that relatively deeper models may require longer training or better normalization between submodules to be effective. We don't see this effect for higher number of channels. Since the best performing model was created with $L=5,C=64$, we decided to use it for the second stage of the ablation study. 

We disabled some modules of our pipeline with Ballé codec with neural networks of size $C=64$ channels and $L=5$ layers. We measured the performance on quality points between 2 and 5. We found out that the CR model alone doesn't produce much gain while RS model alone is reasonably performant. However, best results come when two modules are combined. The desaturation transformation hurts the performance alone but the introduction of RS model on top of this results in bitrate savings. When the full pipeline is activated, we see that every module contributes and we get the best possible performance.

\begin{figure}[!htb]
\begin{subfigure}[b]{0.48\linewidth}
  \centering
  \includegraphics[width=1.0\linewidth]{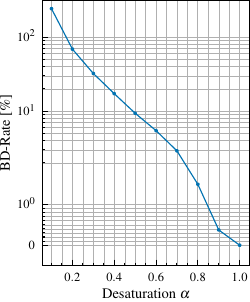}
\end{subfigure}
\hfill
\begin{subfigure}[b]{0.48\linewidth}
  \centering
  \includegraphics[width=1.0\linewidth]{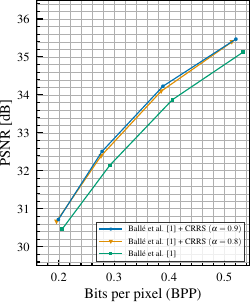}
\end{subfigure}
\caption{Effect of desaturation coefficient $\alpha$ on top of Ballé baseline codec. Left image shows the BD-Rate without the neural network modules and the right figure depicts the Rate-Distortion plot of $\alpha=0.8$ and $\alpha=0.9$ with the neural network modules.}
\label{fig:alphaperf}
\end{figure}

In our experiments, we also tested the effect of desaturation coefficient $\alpha$ on the performance of our pipeline. From Fig.~\ref{fig:alphaperf}, it can be seen that there is an inverse log-linear correlation between the desaturation coefficient and the BD-Rate when no neural networks are activated in the pipeline. We investigated the sudden drop of the curve around $\alpha=0.9$ by training our neural pipeline to see whether it can produce better gain than its $\alpha=0.8$ counterpart. We observed that the model with $\alpha=0.9$ reaches a slightly higher performance gain of 1.1\% compared to the model with $\alpha=0.8$. This shows that the desaturation coefficient can be further optimized for a higher performance. 

\section{Conclusion}
\label{sec:conclusion}

In this paper, we tried to solve the adaptation of LICs to SC data without breaking the codec's self consistency. To do so, we proposed two neural networks, CR and RS models to act as the prefilter and the postfilter. We also demonstrated that introduction of some forward and inverse transformations help with the codec efficiency. Specifically, the desaturation transformation can reduce the bitrate required to compress an image via destroying some information. However, the neural networks are able to recover from this transformation for reasonable magnitude. It can be clearly seen that with a little bit of extra compute cost, it is possible to adapt LICs to unseen domains without retraining the codecs themselves.

While the performance improvement shown here is promising, LICs still require lots of compute power compared to the classical codecs. This creates a barrier of adaptation for standardized LICs and introducing new networks may not be favorable in some use cases. Creating more efficient and faster LICs is research direction with lots of real world applications. Another interesting approach could be in adapting this approach to other special domains such as the medical, microscopy, multispectral or space imaging where more efficient codecs are still in need. These new domains may require new specialized transformations and architectures. Finally, it is also a promising research direction to find optimal transformation coefficients per image to get even more specialized gains.



\bibliographystyle{IEEEbib}
\bibliography{strings,refs}

\begin{thebibliography}{10}

\bibitem{balle2018variational}
Johannes Ball{\'e}, David Minnen, Saurabh Singh, Sung~Jin Hwang, and Nick
  Johnston,
\newblock ``Variational image compression with a scale hyperprior,''
\newblock {\em arXiv preprint arXiv:1802.01436}, 2018.

\bibitem{minnen2018joint}
David Minnen, Johannes Ball{\'e}, and George~D Toderici,
\newblock ``Joint autoregressive and hierarchical priors for learned image
  compression,''
\newblock {\em Advances in neural information processing systems}, vol. 31,
  2018.

\bibitem{cheng2020learned}
Zhengxue Cheng, Heming Sun, Masaru Takeuchi, and Jiro Katto,
\newblock ``Learned image compression with discretized gaussian mixture
  likelihoods and attention modules,''
\newblock in {\em Proceedings of the IEEE/CVF conference on computer vision and
  pattern recognition}, 2020, pp. 7939--7948.

\bibitem{koyuncu2022contextformer}
A.~Burakhan Koyuncu, Han Gao, Atanas Boev, Georgii Gaikov, Elena Alshina, and
  Eckehard Steinbach,
\newblock ``Contextformer: A transformer with spatio-channel attention for
  context modeling in learned image compression,''
\newblock in {\em Computer Vision--ECCV 2022: 17th European Conference, Tel
  Aviv, Israel, October 23--27, 2022, Proceedings, Part XIX}. Springer, 2022,
  pp. 447--463.

\bibitem{koyuncu2023efficient}
A.~Burakhan Koyuncu, Panqi Jia, Atanas Boev, Elena Alshina, and Eckehard
  Steinbach,
\newblock ``Efficient contextformer: Spatio-channel window attention for fast
  context modeling in learned image compression,''
\newblock {\em arXiv preprint arXiv:2306.14287}, 2023.

\bibitem{he2022elic}
Dailan He, Ziming Yang, Weikun Peng, Rui Ma, Hongwei Qin, and Yan Wang,
\newblock ``Elic: Efficient learned image compression with unevenly grouped
  space-channel contextual adaptive coding,''
\newblock in {\em Proceedings of the IEEE/CVF Conference on Computer Vision and
  Pattern Recognition}, 2022, pp. 5718--5727.

\bibitem{liu2023learned}
Jinming Liu, Heming Sun, and Jiro Katto,
\newblock ``Learned image compression with mixed transformer-cnn
  architectures,''
\newblock in {\em Proceedings of the IEEE/CVF Conference on Computer Vision and
  Pattern Recognition}, 2023, pp. 14388--14397.

\bibitem{Wang2022}
Yingbin Wang, Xin Zhao, Xiaozhong Xu, Shan Liu, Zhijun Lei, Mariana Afonso,
  Andrey Norkin, and Thomas Daede,
\newblock ``An open video dataset for screen content coding,''
\newblock in {\em 2022 Picture Coding Symposium (PCS)}. Dec. 2022, IEEE.

\bibitem{hevc}
Gary~J. Sullivan, Jens-Rainer Ohm, Woo-Jin Han, and Thomas Wiegand,
\newblock ``Overview of the high efficiency video coding (hevc) standard,''
\newblock {\em IEEE Transactions on Circuits and Systems for Video Technology},
  vol. 22, no. 12, pp. 1649--1668, 2012.

\bibitem{vvc}
Benjamin Bross, Ye-Kui Wang, Yan Ye, Shan Liu, Jianle Chen, Gary~J. Sullivan,
  and Jens-Rainer Ohm,
\newblock ``Overview of the versatile video coding (vvc) standard and its
  applications,''
\newblock {\em IEEE Transactions on Circuits and Systems for Video Technology},
  vol. 31, no. 10, pp. 3736--3764, 2021.

\bibitem{liu2015overview}
Shan Liu, Xiaozhong Xu, Shawmin Lei, and Kevin Jou,
\newblock ``Overview of hevc extensions on screen content coding,''
\newblock {\em APSIPA Transactions on Signal and Information Processing}, vol.
  4, pp. e10, 2015.

\bibitem{overview_recent_video_scc}
Xiaozhong Xu and Shan Liu,
\newblock ``Overview of screen content coding in recently developed video
  coding standards,''
\newblock {\em IEEE Transactions on Circuits and Systems for Video Technology},
  vol. 32, no. 2, pp. 839--852, 2022.

\bibitem{balle2016end}
Johannes Ball{\'e}, Valero Laparra, and Eero~P Simoncelli,
\newblock ``End-to-end optimized image compression,''
\newblock {\em arXiv preprint arXiv:1611.01704}, 2016.

\bibitem{comcnn}
Wen Tao, Feng Jiang, Shengping Zhang, Jie Ren, Wuzhen Shi, Wangmeng Zuo, Xun
  Guo, and Debin Zhao,
\newblock ``An end-to-end compression framework based on convolutional neural
  networks,''
\newblock in {\em 2017 Data Compression Conference (DCC)}, 2017, pp. 463--463.

\bibitem{sandwich1}
Onur~G. Guleryuz, Philip~A. Chou, Hugues Hoppe, Danhang Tang, Ruofei Du, Philip
  Davidson, and Sean Fanello,
\newblock ``Sandwiched image compression: Wrapping neural networks around a
  standard codec,''
\newblock in {\em 2021 IEEE International Conference on Image Processing
  (ICIP)}, 2021, pp. 3757--3761.

\bibitem{sandwich2}
Onur~G. Guleryuz, Philip~A. Chou, Hugues Hoppe, Danhang~"Danny" Tang, Ruofei
  Du, Philip Davidson, and Sean Fanello,
\newblock ``Sandwiched image compression: Increasing the resolution and dynamic
  range of standard codecs,''
\newblock in {\em 2022 Picture Coding Symposium (PCS)}, 2022,
\newblock Best Paper Finalist.

\bibitem{Penrose1955}
Roger Penrose,
\newblock ``A generalized inverse for matrices,''
\newblock {\em Mathematical Proceedings of the Cambridge Philosophical
  Society}, vol. 51, no. 3, pp. 406–413, July 1955.

\bibitem{Lloyd1982}
Stuart Lloyd,
\newblock ``Least squares quantization in pcm,''
\newblock {\em IEEE Transactions on Information Theory}, vol. 28, no. 2, pp.
  129–137, Mar. 1982.

\bibitem{He2016}
Kaiming He, Xiangyu Zhang, Shaoqing Ren, and Jian Sun,
\newblock ``Deep residual learning for image recognition,''
\newblock in {\em 2016 IEEE Conference on Computer Vision and Pattern
  Recognition (CVPR)}. June 2016, IEEE.

\bibitem{effnet}
Mingxing Tan and Quoc Le,
\newblock ``{E}fficient{N}et: Rethinking model scaling for convolutional neural
  networks,''
\newblock in {\em Proceedings of the 36th International Conference on Machine
  Learning}, Kamalika Chaudhuri and Ruslan Salakhutdinov, Eds. 09--15 Jun 2019,
  vol.~97 of {\em Proceedings of Machine Learning Research}, pp. 6105--6114,
  PMLR.

\bibitem{shi2016real}
Wenzhe Shi, Jose Caballero, Ferenc Husz{\'a}r, Johannes Totz, Andrew~P Aitken,
  Rob Bishop, Daniel Rueckert, and Zehan Wang,
\newblock ``Real-time single image and video super-resolution using an
  efficient sub-pixel convolutional neural network,''
\newblock in {\em Proceedings of the IEEE conference on computer vision and
  pattern recognition}, 2016, pp. 1874--1883.

\bibitem{ascenso2023jpeg}
Jo{\~a}o Ascenso, Elena Alshina, and Touradj Ebrahimi,
\newblock ``The jpeg ai standard: Providing efficient human and machine visual
  data consumption,''
\newblock {\em IEEE MultiMedia}, vol. 30, no. 1, pp. 100--111, 2023.

\bibitem{kingma2014adam}
Diederik~P Kingma and Jimmy Ba,
\newblock ``Adam: A method for stochastic optimization,''
\newblock {\em arXiv preprint arXiv:1412.6980}, 2014.

\bibitem{begaint2020compressai}
Jean B{\'e}gaint, Fabien Racap{\'e}, Simon Feltman, and Akshay Pushparaja,
\newblock ``Compressai: a pytorch library and evaluation platform for
  end-to-end compression research,''
\newblock {\em arXiv preprint arXiv:2011.03029}, 2020.

\bibitem{Bjontegaard2001CalculationOA}
Gisle Bj{\o}ntegaard,
\newblock ``Calculation of average psnr differences between rd-curves,''
\newblock 2001.

\bibitem{herglotz2022beyond}
Christian Herglotz, Matthias Kr{\"a}nzler, Ruben Mons, and Andr{\'e} Kaup,
\newblock ``Beyond bj{\o}ntegaard: Limits of video compression performance
  comparisons,''
\newblock in {\em 2022 IEEE International Conference on Image Processing
  (ICIP)}. IEEE, 2022, pp. 46--50.

\end{thebibliography}

\end{document}